\documentclass[fleqn,usenatbib,useAMS]{mnras}

\usepackage{graphicx}	
\usepackage{amsmath}	
\usepackage{amssymb}	
\usepackage{multicol}        
\usepackage{pdflscape}	
\usepackage{cancel}     
\usepackage{subcaption}
\usepackage{caption}
\setcitestyle{notesep={ }}

\newcommand{\red}[1]{\textcolor{black}{#1}}

\newcommand{\ahod}{\textsc{AbacusHOD}}

\usepackage[T1]{fontenc}
\usepackage{ae,aecompl}

\usepackage{newtxtext,newtxmath}

\title[2D $k$NN]{2D $k$-th nearest neighbor statistics: a highly informative probe of galaxy clustering}

\author[Yuan et al]{
Sihan Yuan\thanks{E-mail: sihany@stanford.edu},
Alvaro Zamora, 
and Tom Abel
\\
Kavli Institute for Particle Astrophysics and Cosmology, Stanford University, 452 Lomita Mall, Stanford, CA 94305, USA\\
Department of Physics, Stanford University, 382 Via Pueblo Mall, Stanford, CA 94305, USA\\
SLAC National Accelerator Laboratory, 2575 Sand Hill Road, Menlo Park, CA  94025, USA
}

\date{Accepted XXX. Received YYY; in original form ZZZ}

\pubyear{2022}

\begin{document}
\label{firstpage}
\pagerange{\pageref{firstpage}--\pageref{lastpage}}
\maketitle

\begin{abstract}
\red{Beyond standard summary statistics are necessary to summarize the rich information on non-linear scales in the era of precision galaxy clustering measurements. For the first time, we introduce the 2D $k$-th nearest neighbor ($k$NN) statistics as a summary statistic for discrete galaxy fields. This is a direct generalization of the standard 1D $k$NN by disentangling the projected galaxy distribution from the redshift-space distortion signature along the line-of-sight. We further introduce two different flavors of 2D $k$NNs that trace different aspects of the galaxy field: the standard flavor which tabulates the distances between galaxies and random query points, and a ``DD'' flavor that tabulates the distances between galaxies and galaxies. We showcase the 2D $k$NNs' strong constraining power both through theoretical arguments and by testing on realistic galaxy mocks. Theoretically, we show that 2D $k$NNs are computationally efficient and directly generate other statistics such as the popular 2-point correlation function, voids probability function, and counts-in-cell statistics. 
In a more practical test, we apply the 2D $k$NN statistics to simulated galaxy mocks that fold in a large range of observational realism and recover parameters of the underlying extended halo occupation distribution (HOD) model that includes velocity bias and galaxy assembly bias. 
We find unbiased and significantly tighter constraints on all aspects of the HOD model with the 2D $k$NNs, both compared to the standard 1D $k$NN, and the classical redshift-space 2-point correlation functions. }
\end{abstract}




\section{Introduction}

Current and upcoming large-scale galaxy surveys will collect an extraordinary amount of data that encode rich information in both cosmology and galaxy physics. Spectroscopic surveys such as the Dark Energy Spectroscopic Instrument \citep[DESI;][]{2016DESI}, the Subaru Prime Focus Spectrograph \citep[PFS;][]{2014Takada}, the ESA \textit{Euclid} satellite mission \citep[][]{2011Laureijs} in addition to photometric surveys like the Rubin observatory-LSST \citep{2019Ivezic} are expected to extend our current reach in both volume and depth by an order of magnitude. The volume and precision of the data require a robust modeling framework that can fully utilize the information content of the datasets and derive stringent and unbiased constraints on cosmology and galaxy physics. 

One key requirement of such robust modeling framework is that it maximally captures the information contained in the clustering data down to small scales ($\approx 1-30$Mpc/$h$), which is coincidentally where modern surveys also have their highest signal-to-noise. Existing analyses of galaxy survey data almost exclusively utilize the galaxy 2-point correlation function (2PCF) as the summary statistics of the galaxy clustering. While the 2PCF is powerful in characterizing Gaussian density fields on large scales, it leaves out significant amount of information when applied to small scales, where the non-linear evolution post shell-crossing and the complex physics of galaxy evolution become important. Therefore, the challenge of extracting the full information from small-scale clustering necessarily require the development and analysis of novel statistics beyond the 2PCF \citep[e.g.][]{2023Paillas, 2022aValogiannis, 2021Banerjee, 2020Hahn, 2020Uhlemann, 2017Yuan, 2017bSlepian}.

Another challenge in modeling small scales is that the standard perturbative models break down because of the complex non-linear evolution. Thus, considerable recent modeling efforts have been devoted to developing large N-body cosmological simulations, which can provide accurate model templates of the dark matter field down to arbitrarily small scales given sufficient computational resources \citep[Recent examples include][]{2021Maksimova, 2019DeRose, 2021Ishiyama}. However, N-body simulations only simulate the gravitational growth of the total matter field, so we still need a robust so-called galaxy--dark matter connection model that ``paints'' galaxies on top of the simulated matter density field in order to generate forward models of the observed galaxy field \citep[See][for a review]{2018Wechsler}. The extra information captured with beyond-2PCF statistics at small scales can be particularly powerful in constraining the galaxy--dark matter connection model, which would not only reveal details of galaxy evolution physics, but also feed back into the cosmology analysis and derive stronger cosmological constraints. 

\red{In this paper, we continue on previous work presented in \citet{2021Banerjee} and introduce observationally motivated generalizations of the $k$-th nearest neighbor ($k$NN) statistics. We demonstrate their constraining power on the galaxy--dark matter connection models through both theoretical arguments and a set of fully realistic mock tests.} For the latter, we adopt the lightcone-based full forward model approach described in \citet{2022cYuan} to conduct parameter recovery tests while accounting for the full range of observational systematics, including redshift evolution and layers of survey incompleteness. We demonstrate that despite the observational systematics, our proposed 2D $k$NN variations yield unbiased yet significantly tighter constraints on galaxy--halo connection than the 2PCF. 

The paper is organized as follows. In section~\ref{sec:mocks}, we describe the basic framework for our models and mock production. In section~\ref{sec:statistics}, we give an overview of the relevant summary statistics, the 2PCF, and \red{the $k$-th nearest neighbor statistics with 2D generalizations}. In section~\ref{sec:recovery}, we describe our fully realistic mock test and compare the model constraints of the 2PCF and the $k$NNs, including a discussion of galaxy assembly bias. In section~\ref{sec:conclude}, we provide additional discussion of our results and draw a few conclusions. 


\section{Simulation and mocks}
\label{sec:mocks}
In this section, we introduce the simulation-based model template used in the ensuing analysis and the construction of our realistic lightcone-based mocks. 

\subsection{\textsc{AbacusSummit} lightcones}
\label{subsec:lightcones}

The \textsc{AbacusSummit} simulation suite \citep[][]{2021Maksimova} is a set of large, high-accuracy cosmological N-body simulations using the \textsc{Abacus} N-body code \citep{2019Garrison, 2021bGarrison}, designed to meet and exceed the Cosmological Simulation Requirements of the Dark Energy Spectroscopic Instrument (DESI) survey \citep{2013arXiv1308.0847L}. \textsc{AbacusSummit} consists of over 150 simulations, containing approximately 60 trillion particles at 97 different cosmologies. 
For this analysis, we use exclusively the ``base'' configuration boxes within the simulation suite, each of which contains $6912^3$ particles within a $(2h^{-1}$Gpc$)^3$ volume, corresponding to a particle mass of $2.1 \times 10^9 h^{-1}M_\odot$. \footnote{For more details, see \url{https://abacussummit.readthedocs.io/en/latest/abacussummit.html}}
The \textsc{AbacusSummit} suite also uses a specialized spherical-overdensity based halo finder known as {\sc CompaSO} \citep{2021Hadzhiyska}.

In addition to periodic boxes, the simulation suite also provides a set of simulation lightcones at fiducial cosmology \citep[][]{2022Hadzhiyska}. The basic algorithm associates the halos from a set of coarsely-spaced snapshots with their positions at the time of light-cone crossing by matching halo particles to on-the-fly light cone particles. The resulting halo catalogs are considered reliable at $M_\mathrm{halo} > 2.1\times 10^{11}h^{-1}M_\odot$. For this analysis, we utilize the 25 base lightcones, which are constructed from the 25 base periodic boxes, with each lightcone covering an octant of the sky ($\sim 5156$ deg$^2$) up to $z\sim 0.8$. 

A suite of lightcones at non-fiducial cosmologies are also being produced and will enable emulator cosmology analysis. 

\subsection{The Halo Occupation Distribution (HOD)}
\label{subsec:model}
The galaxy--halo connection model we use for generating the realistic mocks and for the forward model is known as the Halo Occupation Distribution \citep[HOD; e.g.][]{2005Zheng, 2007bZheng}, which probabilistically populate dark matter halos with galaxies according to a set of halo properties. For a Luminous Red Galaxy (LRG) sample, the HOD is well approximated by a vanilla model given by (originally shown in \citet{2015Kwan}):
\begin{align}
    \bar{n}_{\mathrm{cent}}^{\mathrm{LRG}}(M) & = \frac{f_\mathrm{ic}}{2}\mathrm{erfc} \left[\frac{\log_{10}(M_{\mathrm{cut}}/M)}{\sqrt{2}\sigma}\right], \label{equ:zheng_hod_cent}\\
    \bar{n}_{\mathrm{sat}}^{\mathrm{LRG}}(M) & = \left[\frac{M-\kappa M_{\mathrm{cut}}}{M_1}\right]^{\alpha}\bar{n}_{\mathrm{cent}}^{\mathrm{LRG}}(M),
    \label{equ:zheng_hod_sat}
\end{align}
where the five vanilla parameters characterizing the model are $M_{\mathrm{cut}}, M_1, \sigma, \alpha, \kappa$. $M_{\mathrm{cut}}$ characterizes the minimum halo mass to host a central galaxy. $M_1$ characterizes the typical halo mass that hosts one satellite galaxy. $\sigma$ describes the steepness of the transition from 0 to 1 in the number of central galaxies. $\alpha$ is the power law index on the number of satellite galaxies. $\kappa M_\mathrm{cut}$ gives the minimum halo mass to host a satellite galaxy.
We have added a modulation term $\bar{n}_{\mathrm{cent}}^{\mathrm{LRG}}(M)$ to the satellite occupation function to remove satellites from halos without centrals. We have also included an incompleteness parameter $f_\mathrm{ic}$, which is a downsampling factor controlling the overall number density of the mock galaxies. This parameter is relevant when trying to match the observed mean density of the galaxies in addition to clustering measurements. By definition, $0 < f_\mathrm{ic}\leq 1$.

In addition to determining the number of galaxies per halo, the standard HOD model also dictates the position of velocity of the galaxies. For the central galaxy, its position and velocity are set to be the same as those the halo center, specifically the L2 subhalo center-of-mass for the {\sc CompaSO} halos. For the satellite galaxies, they are randomly assigned to halo particles with uniform weights, each satellite inheriting the position and velocity of its host particle. 

For this paper, we fix two parameters $\sigma$ and $\kappa$ for simplicity. $\kappa$ does not strongly affect clustering and only comes into effect at very small scales. $\sigma$ does affect clustering on 2-halo scales, but it tends to be strongly degenerate with $\log M_\mathrm{cut}$. We omit $\sigma$ in this preliminary analysis for clearer interpretation of the results. 

We also include an additional HOD extension known as velocity bias, which biases the velocities of the central and satellite galaxies. This is shown to to be a necessary ingredient in modeling BOSS LRG redshift-space clustering on small scales \citep[e.g.][]{2015aGuo, 2021bYuan}. Velocity bias has also been identified in hydrodynamical simulations and measured to be consistent with observational constraints \citep[e.g.][]{2022Yuan, 2017Ye}. 

We parametrize velocity bias through two additional HOD parameters: \texttt{$\alpha_\mathrm{vel, c}$} is the central velocity bias parameter, which modulates the peculiar velocity of the central galaxy relative to the halo center. $\alpha_\mathrm{vel, c} = 0$ indicates no central velocity bias, i.e. centrals perfectly track the velocity of halo centers. 
\texttt{$\alpha_\mathrm{vel, s}$} is the satellite velocity bias parameter, which modulates how the satellite galaxy peculiar velocity deviates from that of the local dark matter particle. $\alpha_\mathrm{vel, s} = 1$ indicates no satellite velocity bias, i.e. satellites perfectly track the velocity of their underlying particles. 

For computational efficiency, we adopt the highly optimized \ahod\ implementation, which significantly speeds up the HOD calculation per HOD parameter combination \citep[][]{2021bYuan}. The code is publicly available as a part of the \textsc{abacusutils} package at \url{http://https://github.com/abacusorg/abacusutils}. Example usage can be found at \url{https://abacusutils.readthedocs.io/en/latest/hod.html}. 

To summarize, for this analysis, the HOD is fully parameterized by 6 parameters, $M_{\mathrm{cut}}, M_1, \alpha$, $\alpha_\mathrm{vel, c}$, $\alpha_\mathrm{vel, s}$, and $f_\mathrm{ic}$. Additionally, in section~\ref{subsec:gab}, we further extend this HOD model with galaxy assembly bias.

\subsection{Realistic mocks}

We construct our realistic mocks on 20 base lightcones \texttt{ph005-024} at Planck cosmology. 
To generate galaxies on these lightcones, we start by picking an HOD that roughly matches the properties of the BOSS CMASS sample. The parameters are $\log M_\mathrm{cut} = 12.8$, $\log M_1 = 13.9$, $\sigma = 0.3$, $\alpha = 1.0$, $\kappa = 0.3$, velocity bias parameters $\alpha_\mathrm{vel, c} = 0.2$, $\alpha_\mathrm{vel, s} = 1$, and an completeness parameter $f_\mathrm{ic} = 0.41$ \citep{2021bYuan}. We then follow the procedures laid out in \citet{2022cYuan} to apply layers of survey realism on to the lightcone mocks, including redshift-based incompleteness, survey geometry and various survey masks. \red{We ignore fiber collision for this analysis. However, we do emphasize that we have devised a generic method of correcting for the effect of fiber collision on any summary statistics via probabilistic redshift recoveries, as laid out in section~3 of \citet{2022cYuan}.}

\section{Summary statistics for galaxy clustering}
\label{sec:statistics}
Commonly, we compare the model to the data through a set of compression known as summary statistics. The use of summary statistics drastically reduces the degrees of freedom in the likelihood function and also marginalizes over stochastic components in the data that are hard to model. However, the downside of summary statistics is that they are lossy compressions and do not typically capture the full information of the density field. Thus, it is important to develop summary statistics that maximize both the simplicity of analysis and the information content, especially considering the high cost and tremendous data volumes of modern large-scale structure experiments. 

In this section, we first revisit the standard 2-point correlation function before introducing the $k$-th nearest neighbor statistics ($k$NN) and its 2D generalizations. All visualizations of the summary statistics and their covariance matrices are measured from the mock galaxy catalogs described in the previous section. 

\subsection{2-point correlation function}
The most widely used summary statistics in large-scale structure studies is the 2-point correlation function (2PCF) and its Fourier space counterpart, the power spectrum. In principle, the 2PCF fully summarizes the information content of the Gaussian random field, which is a valid approximation on very large scales but quickly breaks down on smaller scales.
Mathematically, the real space two-point correlation function $\xi(r)$ describes the excess probability above Poisson samples for a galaxy to be located in subvolume $dV_1$ at distance $r$ from another galaxy in subvolume $dV_2$.
\begin{equation}
 dP_{12} = \bar{n}^2 \left[1 + \xi(r)\right] dV_1 dV_2,
\end{equation}
where $\bar{n}$ is the average number density of galaxies over the volume considered \citep{1980Peebles}. 

In redshift space, the Line-of-sight (LoS) positions of galaxies are distorted by their peculiar velocities. This effect is known as redshift-space distortions. However, we can still exploit rotation symmetry around the LoS axis and consider the full shape 2PCF in redshift space as a 2D function $\xi(r_p, r_\pi)$. $r_p$ and $r_\pi$ are the transverse and LoS separations in comoving units. In practice, $\xi(r_p, r_\pi)$ can be computed via the unbiased \citet{1993Landy} estimator:
\begin{equation}
    \xi(r_p, r_\pi) = \frac{DD - 2DR + RR}{RR},
    \label{equ:xi_def}
\end{equation}
where $DD$, $DR$, and $RR$ are the normalized numbers of data-data, data-random, and random-random pair counts in each bin of $(r_p, r_\pi)$. The transverse vs LoS decomposition is particularly convenient as it forms the natural basis that different physical phenomena and systematics operate in. Most significantly, redshift-space distortion, which is the clustering modulation due to velocity dispersion along the LoS only affects clustering along $r_\pi$. While the systematic effects such as fiber collision and survey geometry only affects the observed clustering along the transverse $r_p$ direction. 

While the 2PCF fully captures the information in a Gaussian field, it fails to capture non-gaussian information sourced in variations in primordial inflationary fluctuations and the late time gravitational evolution \citep[e.g.][]{2012Carron, 2003Takada}. On the observational front, the 2PCF is also challenging because it is sensitive to a range of observational systematics, particularly incompleteness due to fiber collision and survey geometry, whose effects propagate through all scales of the 2PCF.  

The 2PCF is also relatively expensive to compute, as the pair finding step in principle scales as the square of the number of points. However, highly-optimized grid-based method to compute the 2PCF have been developed, such as \textsc{Corrfunc} \citep{2020Sinha}, which employs various algorithmic optimizations and hardware acceleration for faster calculations.

\begin{figure}
    \centering
    \hspace*{-0.2cm}
    \includegraphics[width = 3.7in]{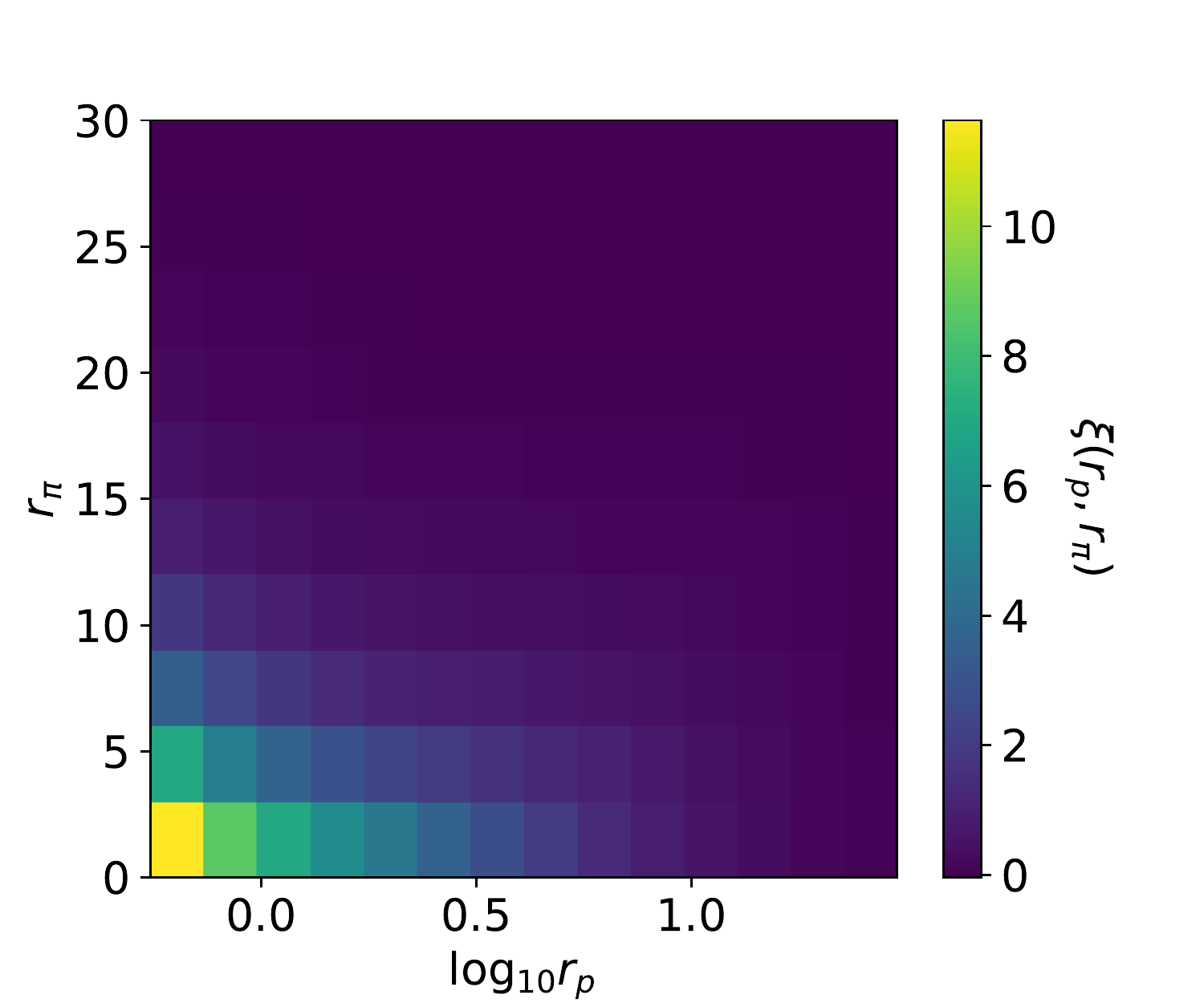}
    \vspace{-0.3cm}
    \caption{The target mock 2PCF $\xi(r_p, r_\pi)$ calculated on the 20 lightcone mocks. We choose the minimum $r_p$ to be 0.5$h^{-1}$Mpc as smaller scales are strongly affected by fiber collision, given the fiber collision recovery scheme that we described in the \citet{2022cYuan}.}
    \label{fig:2pcf_target}
\end{figure}

\begin{figure}
    \centering
    \hspace*{-0.2cm}
    \includegraphics[width = 3.7in]{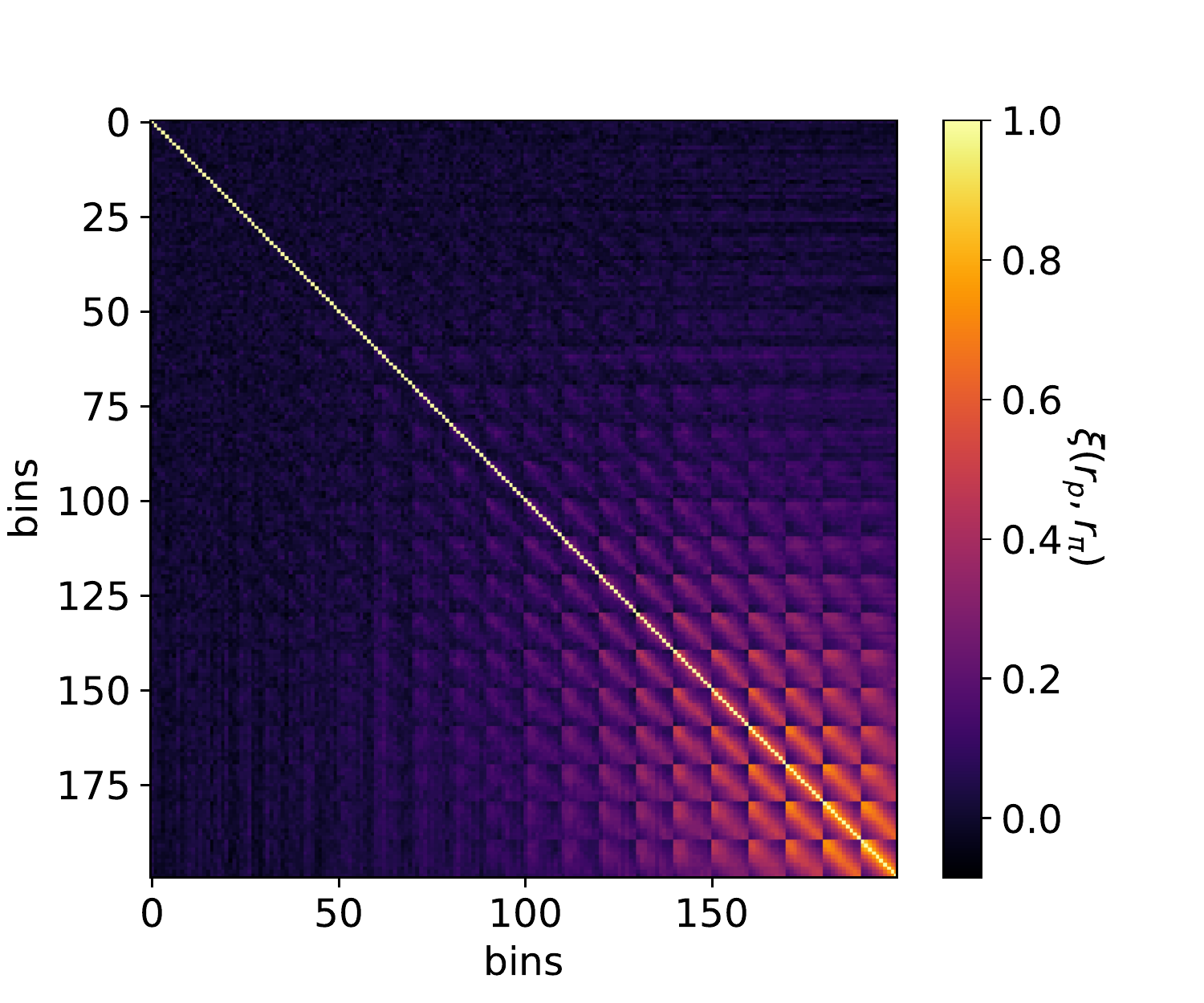}
    \vspace{-0.3cm}
    \caption{The mock correlation matrix of the two-dimensional 2PCF $\xi(r_p. r_\pi)$. The $x$ and $y$ axes denote the bins collapsed to 1D, by concatenating the columns shown in Figure~\ref{fig:2pcf_target}. \red{The covariance matrix is calculated by applying a fiducial HOD to the 1800 \textsc{AbacusSummit} covariance boxes, each of volume $(500h^{-1}$Mpc$)^3$. We refer the readers to section~4 of \citet{2022cYuan} for details.}}
    \label{fig:cov_2pcf}
\end{figure}

For this paper, we use the full-shape redshift-space 2PCF $\xi(r_p, r_\pi)$ for comparison with other statistics and choose the following binning: 14 logarithmic bins in $r_p$ between 0.5$h^{-1}$Mpc and 30$h^{-1}$Mpc and 10 linear bins in $r_\pi$ between 0 and $30h^{-1}$Mpc, for a total of 140 bins. We choose the minimum $r_p$ to be 0.5$h^{-1}$Mpc as smaller scales are strongly affected by fiber collision, given the fiber collision recovery scheme that we described in the \citet{2022cYuan}. We calculate $\xi(r_p, r_\pi)$ in such configuration on a HOD-based galaxy mock resembling CMASS and visualize the result in Figure~\ref{fig:2pcf_target}. We showcase the corresponding covariance matrix in Figure~\ref{fig:cov_2pcf}, where the axes represent the bins flattened by column, with $r_p$ increasingly monotonically in bin number. 

\subsection{1D $k$NNs}
\label{subsec:$k$NNs}

The limits of the 2PCF on small scales necessitates the development of additional summary statistics. One such statistic is the $k$-th nearest neighbor statistics,  or $k$NNs for short \citep[][]{2021Banerjee, 2021bBanerjee, 2021Wang}. The $k$NNs are conceptually straightforward -- measuring the distances to the $k$-th nearest neighbor galaxy in a dataset measured from a set of volume-filling query points, and computing the cumulative distribution of these distances. \cite{2021Banerjee, 2021bBanerjee} showed through Fisher forecasts that $k$NNs are highly informative on cosmology and are sensitive to all orders of $n$-point correlation functions. We offer a brief review in this sub-section.

To construc the 1D $k$NNs, we start with a set of $N_D$ data points in total volume $V_{tot}$. We place a set of $N_R$ random points over the entire volume. Given these random points, we compute the distance from each of these random points to the $k$-th nearest neighbor in the original dataset for a given value of $k$. This can be done very efficiently with the construction of a $KD$-Tree on the data points. We then compute the empirical CDF of distances to the $k$-th nearest neighbor as ${\rm CDF}_{k{\rm NN}}(r)$, i.e. the fraction of random points with distance to the $k$-th nearest neighbor $<r$. This CDF is precisely also the fraction of randomly placed spheres enclosing $> k-1$ points at a given $r$. We can write this down as
\begin{equation}
    {\rm CDF}_{k{\rm NN}}(r) = P_{>k-1|V}|_{V=\frac{4 \pi }{3}r^3}\,.
    \label{equ:pk}
\end{equation}

Because of the nature of CDFs, for large scales very significant changes in the $k$NN-CDF can be of very small magnitude as the statistic asymptotes to 1. Therefore, for visualization purposes it is sometimes beneficial to plot the peaked CDF (${\rm PCDF}$), defined as:
\begin{equation}
    {\rm PCDF(r) = \begin{cases}
        {\rm CDF}(r) \qquad &{\rm CDF}(r) \leq 0.5 \\
        1 - {\rm CDF}(r) \qquad & {\rm CDF}(r) >0.5 \, .
    \end{cases}}
\end{equation}
We refer the reader to \cite{2021Banerjee} for example illustrations of the peaked CDF. The visualizations in this paper are shown in the CDF form. 

For physical intuition, the $k$NN-CDFs are directly related to counts-in-cell statistics. Specifically, from Equation~\ref{equ:pk}, we can write down the counts-in-cell distribution as the difference between two $k$NNs. Specifically, starting from a random point, the probability of finding exactly $k$ neighbors at radius $r$ is
\begin{equation}
    P_{k|V} = {\rm CDF}_{k{\rm NN}}(r) - {\rm CDF}_{(k+1){\rm NN}}(r).
    \label{equ:pkv}
\end{equation}
This shows that the counts-in-cells and $k$NNs at a given radius are equivalent descriptions of the underlying data, and that one
can easily be computed once the other is known. Additionally, $P_{0|V}$ is also known as the void probability function, which characterizes the distribution of voids as a function of void size \citep{1979White}. \red{The fact that both counts-in-cell statistics and void probability functions can be straightforwardly derived from $k$NNs is a strongly indicative of its rich information content. }

\subsection{2D $k$NNs}

\begin{figure*}
    \centering
    \hspace*{-0.4cm}
    \includegraphics[width=\textwidth]{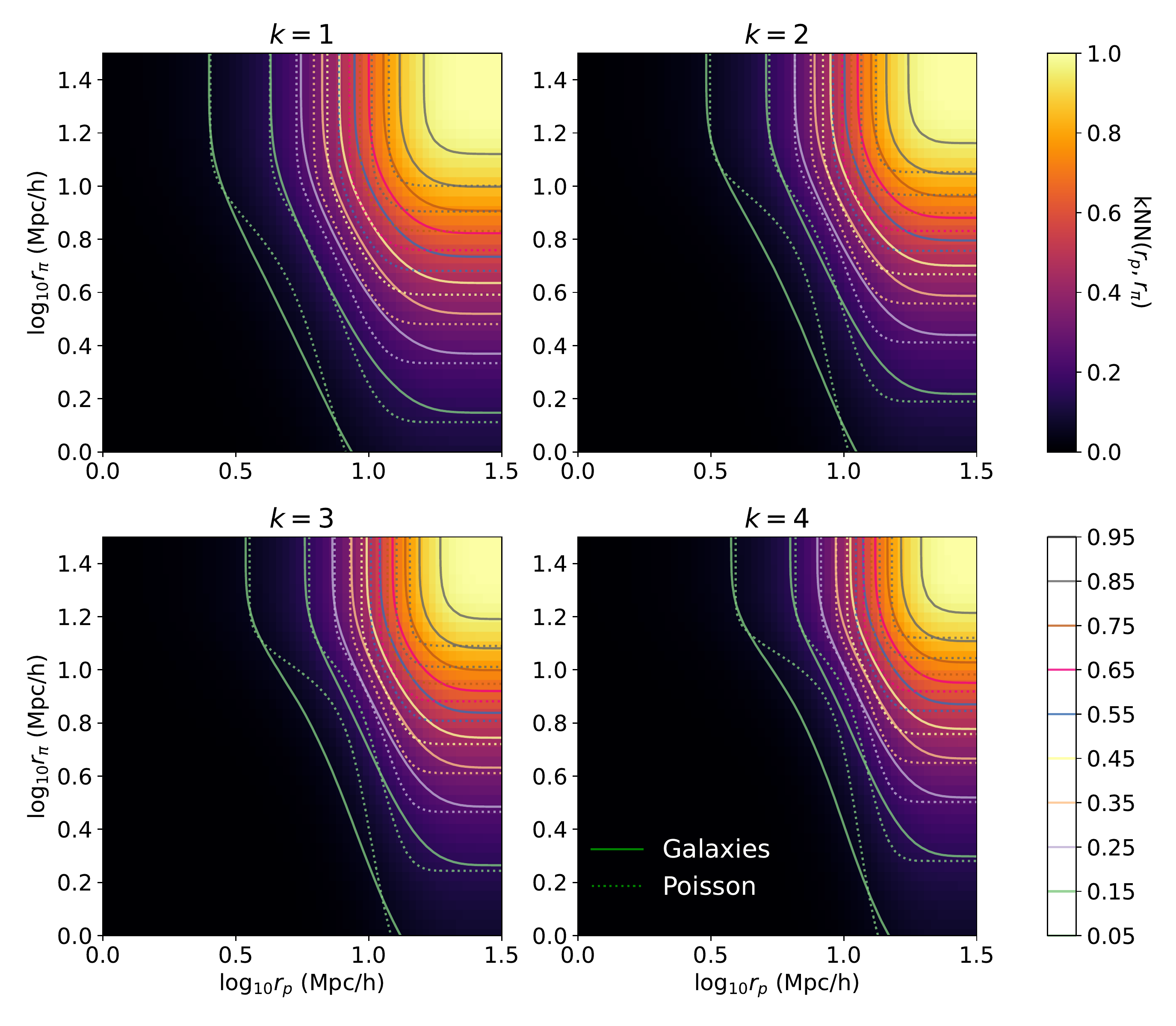}
    \caption{\red{Visualizations of the 2D $k$NN$(r_p, r_\pi)$ as a function of transverse separation $r_p$ and line-of-sight separation $r_\pi$. The 2D heatmap showcases the 2D CDF for $k = 1,2,3,4$. The solid contours highlight the features of the CDF. The dashed contours show the same CDF values for a Poisson random field of the same number of galaxies. The difference in the solid and dashed lines showcase the regions containing the clustering information.} }
    \label{fig:kNN_vis}
\end{figure*}

While we have just described $k$NNs as 1D functions of radial separation $r$, one can also decouple the distance dependency into a LoS component $r_\pi$ and a transverse component $r_p$, analogous to the scale decomposition we introduced for the 2PCF. We speculate that we could gain significantly more information compared to the 1D $k$NN($r$) by disentangling the effects of the redshift-space distortion signatures on the small scales. This point was showcased for the 2PCF in \citet{2021bYuan}.

Thus, we introduce the 2D $k$NN-CDF in the ($r_p, r_\pi$) basis. Specifically, we decompose the distance between each query-data pair $r$ into a $r_p$ and a $r_\pi$ component, and we bin both projections into a 2D histogram. Then we calculate a 2D CDF where each bin accumulates the counts from all bins with smaller $r_p$ and $r_\pi$. Finally, we normalize the cumulative counts by the total number of query points. Conceptually, the 2D $k$NN-CDF is exactly analogous to the default 1D $k$NN-CDF, except we tabulate distances and the cumulative statistics in 2D. For the rest this paper, we denote the default 1D $k$NN-CDF as ``$k$NN($r$)'' and the new 2D $k$NN-CDF as ``$k$NN($r_p, r_\pi$)''. 

\red{For intuition, we plot the first four orders of $k$NN($r_p, r_\pi$) in Figure~\ref{fig:kNN_vis}. In each panel, the 2D heatmap and the solid contours showcase the 2D $k$NN CDF of a mock galaxy sample resembling CMASS. By definition, the CDF monotonically increases from 0 at very small separation to 1 at very large separation. In each panel, the solid contours correspond to CDF values of 0.05, 0.15, 0.25,..., 0.95. The dotted contours showcase the corresponding CDF of an un-clustered Poisson random sample of the same size. The difference between the solid and dotted lines showcase the signatures of clustering. As expected, for larger CDF values, the solid contours reach the same CDF values at larger radii. This is because the query points uniformly sample the entire volume, and thus predominantly occupy voids. As a result, the distance between queries and galaxies can be thought of as a measure of void sizes. Qualitatively, the more clustered the galaxies are, the larger the voids tend to be, and thus the larger distances between queries and their $k$-th nearest galaxy neighbors.}

For this analysis, we set up our $k$NN($r_p, r_\pi$) data vector as the following: we use 8 logarithmic bins along the $r_p$ direction between $0.32h^{-1}$Mpc and $63h^{-1}$Mpc, and 5 logarithmic bins along the $r_\pi$ direction between $1h^{-1}$Mpc and $32h^{-1}$Mpc. We start with $k = 1,2,3,..., 10$ similar to the 1D case. 
\red{The corresponding covariance matrix is shown in Figure~\ref{fig:$k$NNrppi_cov}. The covariance matrix is calculated by applying a fiducial HOD to the 1800 \textsc{AbacusSummit} covariance boxes, each of volume $(500h^{-1}$Mpc$)^3$. We refer the readers to section~4 of \citet{2022cYuan} for details. The relatively coarse sampling of scale along $r_p$ and $r_\pi$ is chosen to control the size of the data vector to be much less than the number of independent mocks available for covariance matrix calculation.} We use the same bin flattening scheme as for $\xi(r_p, r_\pi)$.

\begin{figure*}
    \centering
    \hspace*{-0.4cm}
    \includegraphics[width=\textwidth]{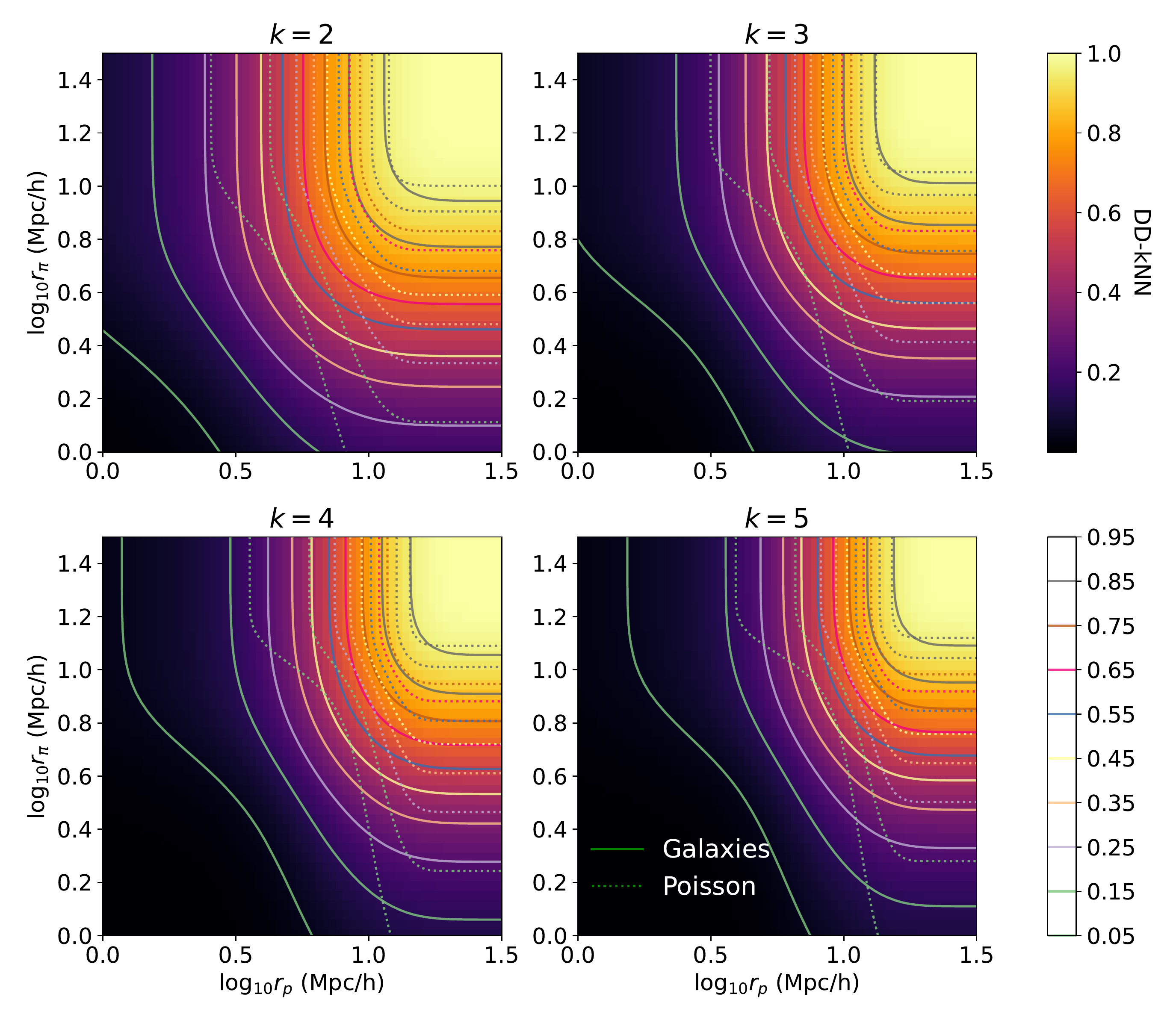}
    \caption{\red{Visualizations of the DD-$k$NN as a function of transverse separation $r_p$ and line-of-sight separation $r_\pi$. The 2D heatmap showcases the 2D CDF for $k = 2,3,4,5$. The solid contours highlight the features of the CDF. The dashed contours show the same CDF values for a Poisson random field of the same number of galaxies. The difference in the solid and dashed lines showcase the regions containing the clustering information.} }
    \label{fig:ddkNN_vis}
\end{figure*}




\begin{figure}
    \centering
    \hspace*{-0.6cm}
    \includegraphics[width = 3.7in]{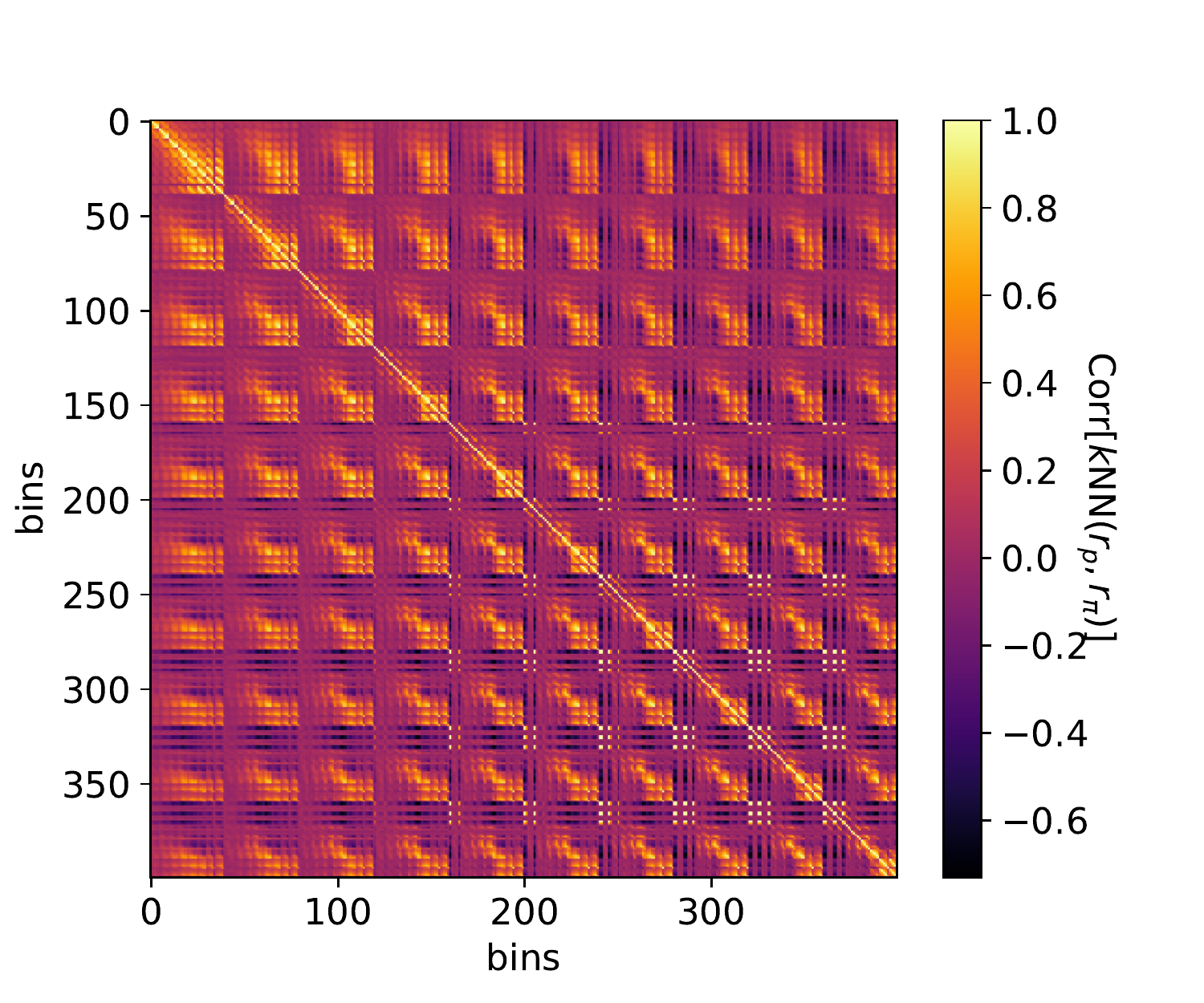}
    \vspace{-0.3cm}
    \caption{The mock correlation matrix of the two-dimensional $k$NN$(r_p. r_\pi)$ for $k = 1,2,3,..., 10$. The $x$ and $y$ axes denote the bins collapsed to 1D, with $r_p$ increasingly monotonically in bin number. \red{The covariance matrix is calculated by applying a fiducial HOD to the 1800 \textsc{AbacusSummit} covariance boxes, each of volume $(500h^{-1}$Mpc$)^3$. We refer the readers to section~4 of \citet{2022cYuan} for details.}}
    \label{fig:$k$NNrppi_cov}
\end{figure}

\subsection{2D Data-data $k$NNs}

We introduce an additional flavor of 2D $k$NN in what we call the data-data $k$NN, or DD-$k$NN. In the standard $k$NN formalism, the distances are measured between data points and query points. However, one can also measure such distances between data points and other data points, analogous to the DD pair counts in correlation statistics. For this reason, the DD-$k$NN is closely related to the 2-point correlation functions and contains a different set of information than the standard $k$NN statistics. Specifically, in the standard $k$NN, the query points are drawn randomly and thus mostly sample the under-dense regions of the density field, but in DD-$k$NN, the query points are the same as data points, thus the DD-$k$NN tracks the over-dense regions, similar to the 2-point correlation function. Because of this, we speculate that the DD-$k$NN is likely more sensitive to small-scale effects and 1-halo scale physics. \red{Another way to think about the difference between the standard $k$NN and the DD-$k$NN is that the standard $k$NN tracers the density field, and is qualitatively a 1-point statistics, whereas the DD-$k$NN traces the clustering, and is thus more akin to a 2-point clustering statistics. In fact, we show in the next paragraphs that the DD-$k$NN exactly derives the 2-point correlation function.} In this paper, we use ``DD-$k$NN'' to specifically refer to the 2D $k$NN($r_p, r_\pi$) but with data-data pairs instead of query-data pairs. 

\red{Figure~\ref{fig:ddkNN_vis} visualizes the DD-$k$NN for $k = 2,3,4,5$. Note that in our definition each galaxy's nearest galaxy neighbor is itself, so DD-$1$NN = 1. Compared to the standard $k$NN($r_p, r_\pi$), the DD-$k$NN CDF reaches the same values at a much smaller separation, consistent with the notion that DD-$k$NN traces high density regions. The contours for an un-clustered Poisson sample also show more dramatic departures from the galaxy contours, suggesting that the DD-$k$NN is highly sensitive to clustering. }

Now we demonstrate that the 2PCF directly derives from the DD-$k$NN. We can think of the 2PCF as the mean overdensity of galaxies around other galaxies, and one can derive this mean overdensity by also integrating over the the PDF derivatives of the DD-$k$NN over all $k$s. Mathematically, we can first write down the 2D PDFs similar to Equation~\ref{equ:pk}
\begin{equation}
    P_{k|\leq r_p, \leq r_\pi} = {\rm CDF}_{k{\rm NN}}(r_p, r_\pi) - {\rm CDF}_{(k+1){\rm NN}}(r_p, r_\pi).
    \label{equ:pk2d}
\end{equation}
$P_{k|\leq r_p, \leq r_\pi}$ gives the probability of finding exactly $k$ galaxies at or within the cylinder defined by radius $r_p$ and height $2r_\pi$ around a given galaxy. Then we can differentiate $P_{k|\leq r_p, \leq r_\pi}$ against both $r_p$ and $r_\pi$ to get the probability of finding the $k$-th galaxy neighbor at ($r_p$ , $r_\pi$) separation away from a given galaxy,
 \begin{equation}
     P_{k|r_p, r_\pi} = \frac{\partial^2P_{k|\leq r_p, \leq r_\pi}}{\partial r_p\partial r_\pi}.
 \end{equation}
 Finaly, we can now express the 2PCF, or the mean overdensity profile of galaxies around a given galaxy as,
 \begin{equation}
     \xi(r_p, r_\pi) = \frac{\sum_{k = 1}^{\infty}kP_{k|r_p, r_\pi}}{DR}-1.
 \end{equation}
 Here we have simply written the denominator as $DR$, which denotes the expected number of galaxies expected at the same separation if the galaxy field is unclustered. One can analytically compute $DR$ as mean density multiplied by the volume of the ($r_p, r_\pi$) voxel. 
 We also point out that while we have adopted the ($r_p, r_\pi$) basis for this derivation, this connection between DD-$k$NN to 2PCF is generic to any scale basis.

\begin{figure}
    \centering
    \hspace*{-0.6cm}
    \includegraphics[width = 3.7in]{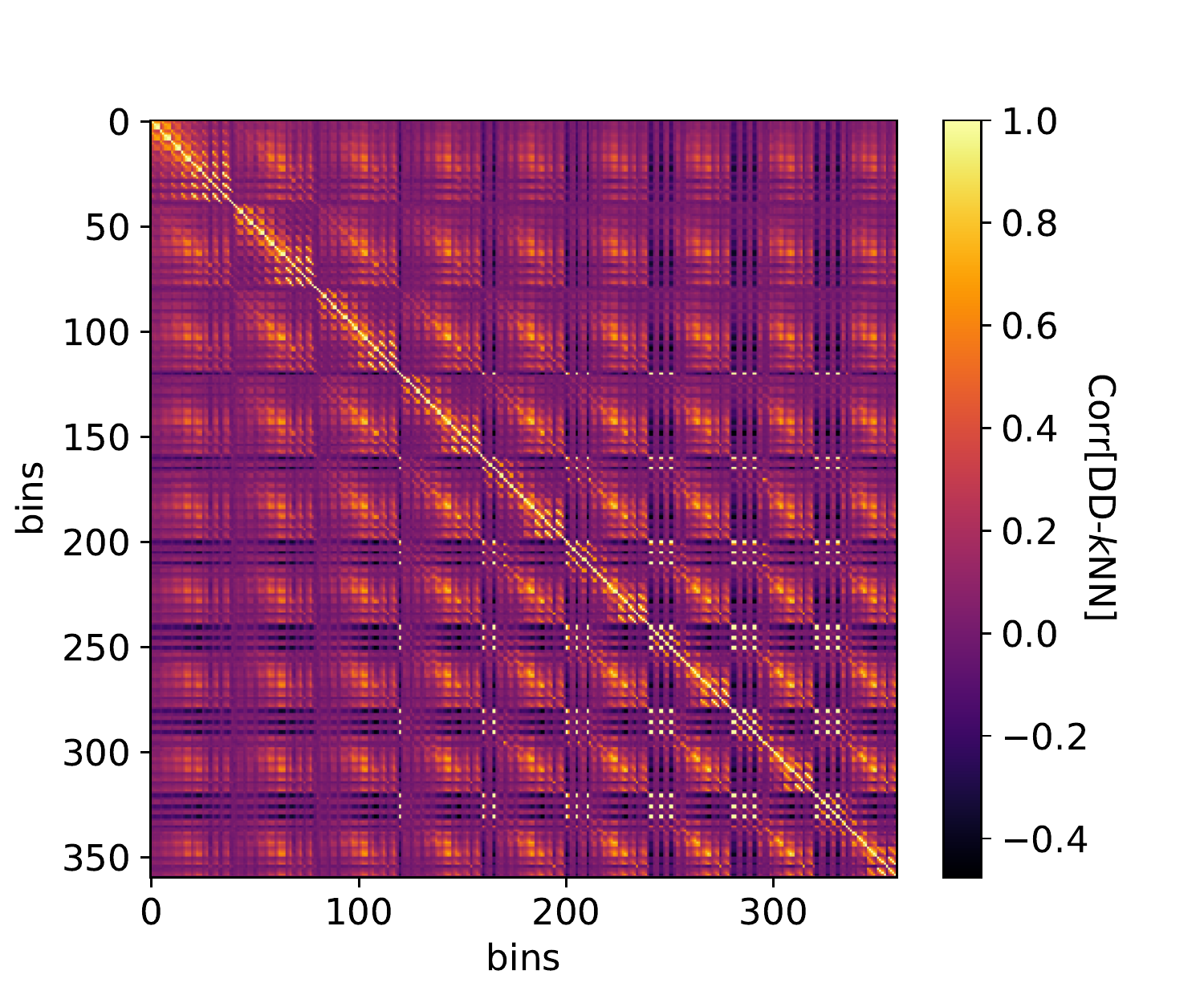}
    \vspace{-0.3cm}
    \caption{The mock correlation matrix of the two-dimensional DD-$k$NN$(r_p. r_\pi)$ for $k = 2,3,..., 10$. The $x$ and $y$ axes denote the bins collapsed to 1D, with $r_p$ increasingly monotonically in bin number. }
    \label{fig:DD$k$NNrppi_cov}
\end{figure}

For this analysis, we set up the DD-$k$NN($r_p, r_\pi$) data vector as follows: we use 8 logarithmic bins along the $r_p$ direction between $0.63h^{-1}$Mpc and $63h^{-1}$Mpc, and 5 logarithmic bins along the $r_\pi$ direction between $0.5h^{-1}$Mpc and $32h^{-1}$Mpc. We include $k$ orders $k = 2,3,..., 10$. Compared to the regular $k$NN($r_p, r_\pi$), we extend the DD-$k$NN down to smaller scales because we expect it to better capture the high density regions and thus show signal down to smaller scales. The corresponding covariance matrix is shown in Figure~\ref{fig:DD$k$NNrppi_cov}.

 \red{Finally, we highlight that despite that the standard query-data $k$NN and the data-data $k$NN are phrased and computed in similar ways, the two statistics are likely sensitive to different scales and trace different moments of the over-density field. We expect the two statistics to be highly complimentary and result in different model degeneracies. We demonstrate this point in the following section.}
 
\subsection{Computational cost}
One of the key advantages of $k$NN and other density-style statistics over higher-order correlation functions is low computational cost. While the cost of $n$-point correlation functions in principle scales with $N^n$, where $N$ is the number of galaxies in the sample, $k$NNs only scales as $N\log N$. In our analysis, we indeed find the $k$NN calculation to be comparable to a highly optimized $N\log N$ 2PCF calculator (\textsc{Corrfunc} \citep{2020Sinha}). 

\red{To further accelerate calculation of $k$NN in both 1D and 2D, we created a publicly available $k$NN library \textsc{fnntw} which achieves high performance by offering parallel builds and a specialized query for the 2D $k$NN. The code is written in \textsc{Rust}\footnote{\url{https://crates.io/crates/fnntw/}} and bindings are made available for \textsc{Python}\footnote{\url{https://pypi.org/project/pyfnntw/}}. Parallelism is achieved in the builds through two avenues not present in other \textsc{Python} $kD$-Tree libraries. Firstly, we note that every subtree of a $kD$-Tree is an independent $kD$-Tree which can be built on a separate thread. The library offers a build method which parallelizes the builds for a subset of these subtrees. Secondly, on a serial build that uses a selection algorithm to find the median and partition the space, we find that approximately 95\% of the run time is spent on the selection algorithm. As such, while the subtree size is above an empirically determined threshold we use a parallel median-of-medians algorithm to find an approximate median with which to partition the space. This shortens the build time of large trees without degrading query performance. For 2D queries, \textsc{fnntw} provides a custom query method which returns the line-of-sight and parallel components, adding a negligible cost over a query that only returns the Euclidean distance. We refer the readers to the linked website for performance benchmarks against other popular $kD$-Tree codes. }

\red{For a reference performance benchmark for cosmology, we populate a $2000h^{-1}$Mpc cubic box with 4 million galaxies with a CMASS-like HOD and also 1 million random query points. The workstation used for this test has dual Intel Xeon Gold 5218 chips clocked at 2.3 GHz for a total of 32 physical cores and 256 GB DDR4-2666 RAM. Using 32 threads, the 2PCF $\xi(r_p, \pi)$ calculation with \textsc{Corrfunc} took 0.44 seconds. In comparison, the DD-$k$NN calculation including $k = 2,3,...,10$ took 1.4 seconds. The standard $k$NN($r_p, r_\pi$) calculation for $k=1,2,...,9$ took 1.7 seconds. While the 2PCF code is faster in this specific setup, we note that \textsc{Corrfunc} is grid-based and thus scales very well at large sample sizes. If we decrease the sample size to 200,000, the tree-based DD-$k$NN compute time drops to 0.11 seconds, whereas the $\xi(r_p, \pi)$ times drops only to 0.15 seconds. This suggests that when confronted with large data volumes with upcoming surveys, it is likely beneficial to develop a grid-based $k$NN code. }

\red{Nevertheless, it is impressive that our computational performance with $k$NN is already comparable to the highly developed 2PCF codes. More importantly, $k$NN calculation scales cheaply in $k$. For example, going from the first 10 $k$s to the first 20 $k$s only increase the compute time by $50\%$. Thus, $k$NNs represent a very cheap way of incorporating high-order clustering information, especially when compared to directly computing $n$-point correlation functions. }


\section{HOD recovery on realistic lightcone mocks}
\label{sec:recovery}

In this section, we test and compare the constraining power of the summary statistics described in section~\ref{sec:statistics} by recovering the underlying HOD parameters on a realistic galaxy mock. We first describe the construction of the lightcone mock, then we describe the modeling pipeline before we present the recovered HOD posterior. 

\subsection{Mock data setup}

We first set up the mock target data vectors using the realistic lightcone-based mocks described in section~\ref{sec:mocks}. The set up is also identical to that of \citet{2022cYuan}, where we describe the layers of realism applied to the mock in great detail. 

For the 2PCF, we use the full 2D decomposition $\xi(r_p, r_\pi)$ as our summary statistic. Specifically, we choose 14 logarithmic bins between 0.5$h^{-1}$Mpc and 30$h^{-1}$Mpc along the transverse direction, and 10 linear bins between 0 and $30h^{-1}$Mpc along the LOS direction, for a total of 140 bins. We visualize the mock target $\xi(r_p, r_\pi)$ in Figure~\ref{fig:2pcf_target}.

For the 1D $k$NN($r$), we use the first 10 orders, $k = 1,2,3,..., 10$. For each $k$, we sample the CDF at 50 linearly spaced scales between $r_\mathrm{min} = 0.1h^{-1}$Mpc and $r_\mathrm{max}=20h^{-1}$Mpc. We further remove scales where the CDF is less than 0.1 or greater than 0.9 as these points tend to highly covariant and lead to very poorly behaved covariance matrices. As a result, we end up with 219 points across 10 $k$s. We refer the readers to Figure~5 and Figure~7 of \citet{2022cYuan} for visualizations of $k$NN($r$).

For the $k$NN($r_p, r_\pi$), we use 8 logarithmic bins along the $r_p$ direction between $0.63h^{-1}$Mpc and $63h^{-1}$Mpc, and 5 logarithmic bins along the $r_\pi$ direction between $1h^{-1}$Mpc and $32h^{-1}$Mpc. We start with $k = 1,2,3,..., 10$ similar to the 1D case. 
Then, similar to the 1D case, we also remove bins where the CDF is less than 0.05 or greater than 0.95, where the variation is noise-dominated. 156 bins remain across the 10 $k$s. We adopt the same scheme for the DD-$k$NN except we skip $k = 1$.

\subsection{Analysis pipeline}

To recover the underlying HOD parameters from the summary statistics computed on the target mocks, we utilize the 5 remaining lightcones, phase \texttt{ph000-004}. For each model evaluation, we propose a set of HOD parameters from a flat prior, populate the 5 lightcones (\texttt{ph000-004}) with the proposed HOD, and then apply the systematics effects, including redshift selection, survey window and masks. Finally, we compute the summary statistics averaged over the 5 lightcones, which we compare with the target summary statistics and calculate likelihoods using the aforementioned covariance matrix. For this analysis, we adopt a Gaussian likelihood function that accounts for both the desired summary statistics and also the average density. Specifically, 
\begin{align}
    \log L = &  -\frac{1}{2}(x_\mathrm{proposed} - x_\mathrm{target})^T \boldsymbol{C}^{-1}(x_\mathrm{proposed} - x_\mathrm{target}) \nonumber\\
    & - \frac{1}{2}\frac{(\bar{n}-\bar{n}_\mathrm{target})^2}{\sigma_n^2}
\end{align}
where $x$ is the desired summary statistic, $\boldsymbol{C}$ is the covariance matrix, and $\bar{n}$ is the mean number density. $\sigma_n$ is the uncertainty on the measured mean number density. For a CMASS-like sample, we quote $\sigma_n = 5\%$.

We sample the parameter posteriors with the \textsc{dynesty} nested sampler \citep{2018Speagle, 2019Speagle}. We impose flat priors bounded with an ellipsoid for all parameters. The ellipsoid is constructed as the minimum-volume ellipsoid that envelopes all training points. We initiate each nested sampling chain with 2000 live points and a stopping criterion of $d\log\mathcal{Z} = 0.01$, where $\mathcal{Z}$ is the evidence. As expected, we achieve excellent fits for both summary statistics, with best-fit $\chi^2$/d.o.f $<1$. 

To accelerate the sampling, we substitute the likelihood evaluation with a fast emulator model. Specifically, we stop the explicit likelihood evaluation after 40,000 likelihood calls and use the samples to train a neural network model, which we then replace the explicit likelihood evaluations with in order to continue sampling. We validate and test our trained model to ensure that the predictions are unbiased and the errors are insignificant compared to other systematic errors. For all three statistics, we arrive at $~30\%$ emulation error compared to the expected sample variance, or $10\%$ increase in amplitude to the covariance matrix. While insignificant, we include this additional error in our analysis. For details of this procedure, we refer the readers to section~4 of the companion paper \citet{2022cYuan}.

\begin{figure*}
    \centering
    \hspace*{-1cm}
    \includegraphics[width = 8in]{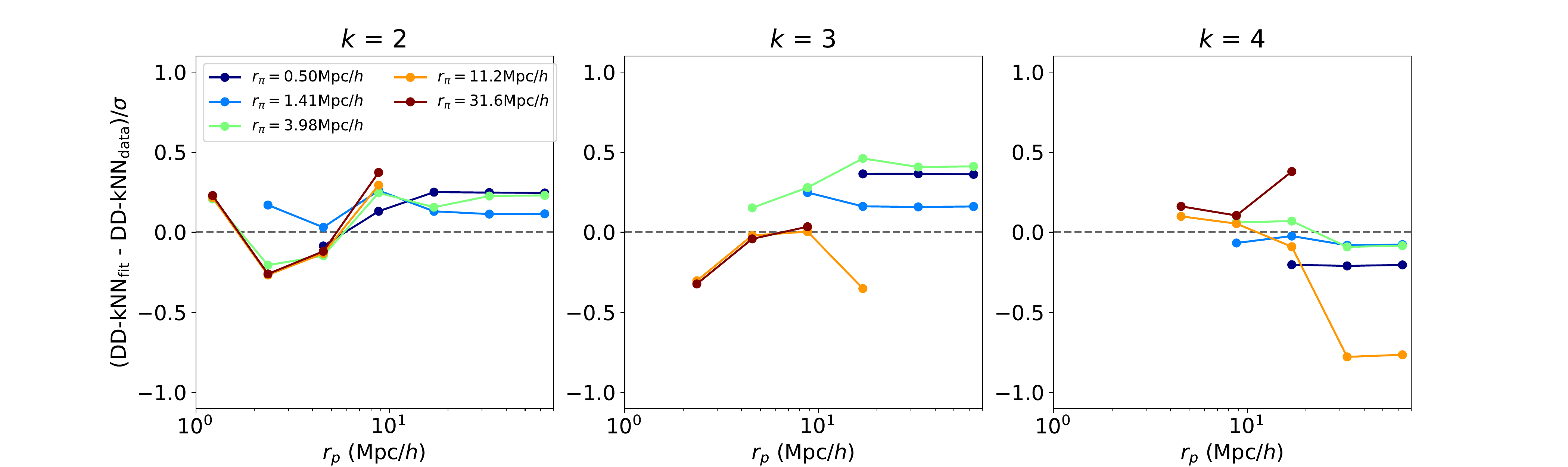}
    \vspace{-0.3cm}
    \caption{The best-fit DD-$k$NN($r_p, r_\pi$) statistic compared to the truth. The $y$-axis shows the difference between the best fit and the mock data vector, normalized by CMASS error bar. The three panels refer to $k = 2, 3, 4$, respectively. Each color refers to a $r_\pi$ bin. The segmentation is due to the bin cuts we placed to remove the tails of the CDF. We see that that we achieve a very good fit. }
    \label{fig:bestfit_kNNs}
\end{figure*}

Figure~\ref{fig:bestfit_kNNs} visualizes the best-fit DD-$k$NN($r_p, r_\pi$) statistic compared to the true mock data vector. Specifically, the $y$-axis shows the relative difference between the best-fit and the mock data normalized by the expected error bars. The panels show different $k$s while the colors represent different $r_\pi$ bins. We see that the best fit is within 1$\sigma$ of the data vector across all bins. The $\sigma$ refers to the expected error bar normalized to the CMASS volume. We omit showing the best-fits of the other data vectors for brevity. 

\subsection{Parameter recovery}

\begin{figure*}
    \centering
    \hspace*{-0.6cm}
    \includegraphics[width = 6in]{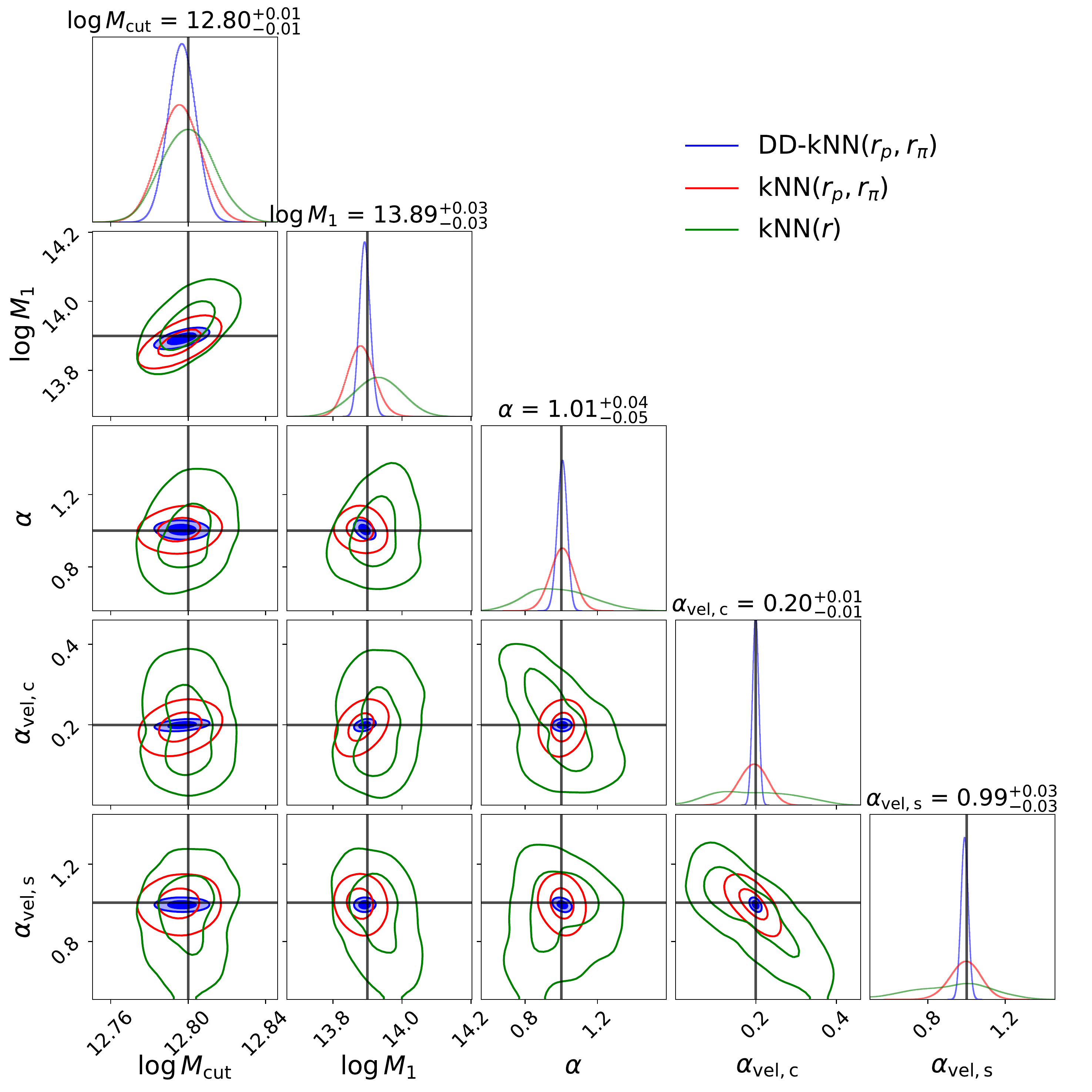}
    \vspace{-0.0cm}
    \caption{The HOD posterior recovery with 3 different flavors of $k$NNs. We only show the $1\sigma$ and $2\sigma$ contours. The black lines indicate the underlying truth. The marginalized constraints shown at the top of the 1D PDFs indicate the mean and the $2\sigma$ constraints of the DD-$k$NN($r_p, r_\pi$) statistic. All three $k$NNs are tabulated from $k = 1$ to $k = 10$, with scale range $0.5$-$30h^{-1}$Mpc. }
    \label{fig:corner_kNNs}
\end{figure*}

\begin{table*}
\centering
\begin{tabular}{lccccc}
\hline
\hline
Parameter     & truth & 1D $k$NN($r$) ($95\%$C.L.) & $k$NN$(r_p, r_\pi)$ ($95\%$C.L.) & DD-$k$NN$(r_p, r_\pi)$ ($95\%$C.L.) & $\xi(r_p, r_\pi)$ ($95\%$C.L.)\\
\hline
$\log_{10}{M_\mathrm{cut}}$  & 12.8 & $12.800\pm 0.024$ & $12.796\pm 0.019$ & $12.797\pm 0.013$ & $12.79\pm 0.02$\\
$\log_{10}{M_{1}}$           & 13.9 & $12.93\pm 0.12$   & $13.88\pm 0.06$   & $13.89\pm 0.03$   & $13.89\pm 0.07$\\
$\alpha$                     & 1.0  & $1.00\pm 0.25$    & $1.00\pm 0.12$    & $0.99\pm 0.05$    & $1.03\pm 0.16$\\
$\alpha_\mathrm{vel, c}$     & 0.2  & $0.21\pm 0.18$    & $0.19\pm 0.06$    & $0.200\pm 0.014$  & $0.20\pm 0.03$\\
$\alpha_\mathrm{vel, s}$     & 1.0  & $0.93\pm 0.36$    & $0.99\pm 0.14$    & $1.00\pm 0.04$    & $0.96\pm 0.08$\\
\hline
$\mathrm{FoM}$                & & 4.2    & 11.4  & 32.0  & 15.5 \\

\hline
\end{tabular}

\caption{The posterior constraints of the 6 HOD model parameters recovered from the 2PCF $w_p$ and variations of the $k$NNs. The bottom row indicates the Figure of Merit (FoM) of the four summary statistics (see Equation~\ref{equ:fom}). }
\label{tab:constraints}
\end{table*}

\begin{figure*}
    \centering
    \hspace*{-0.6cm}
    \includegraphics[width = 6in]{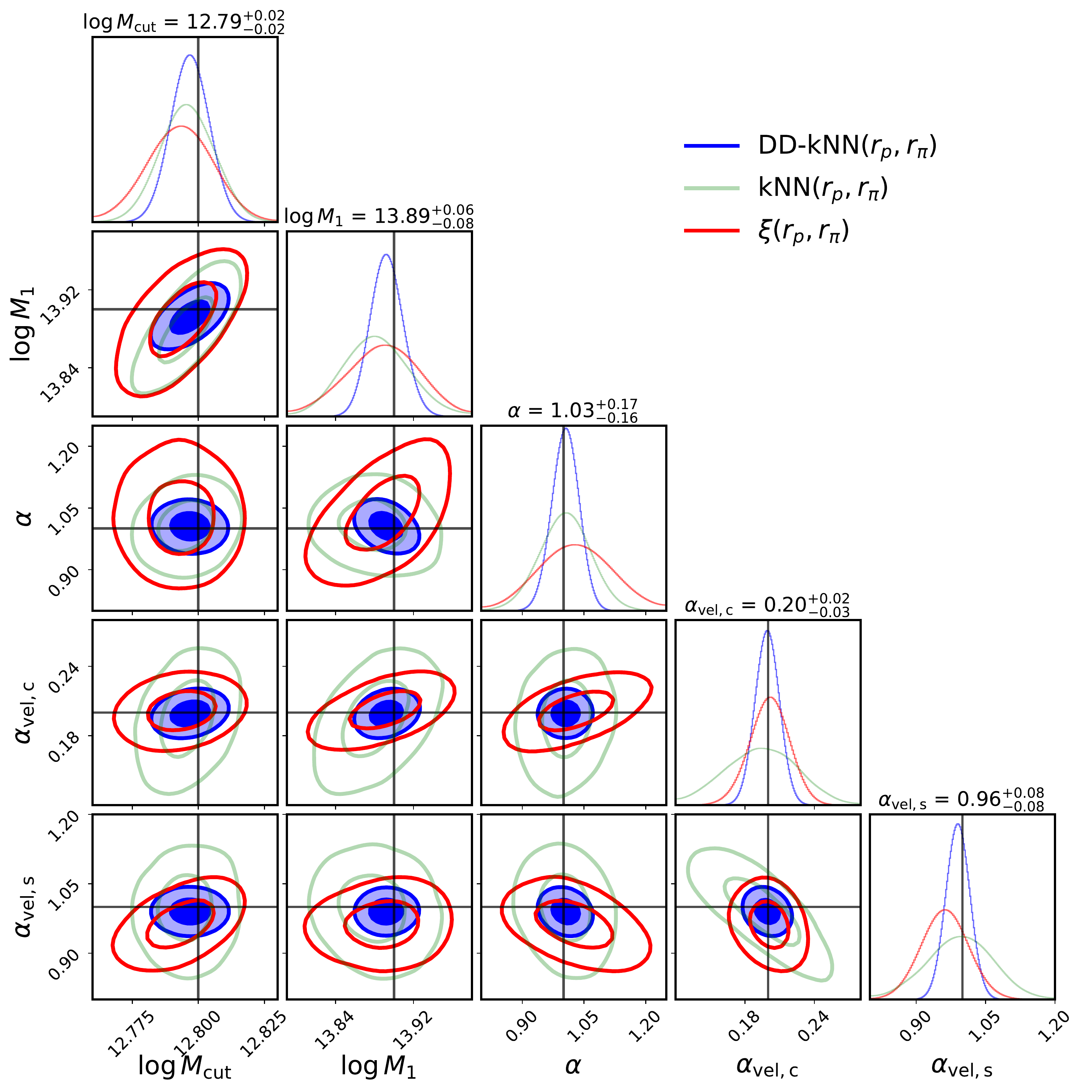}
    \vspace{-0.0cm}
    \caption{The HOD posterior recovery with the redshift-space 2PCF compared with the 2D $k$NNs. We show the 1 and 2$\sigma$ contours. The black lines indicate the underlying truth. The marginalized constraints shown at the top of the 1D PDFs indicate the $2\sigma$ constraints of $\xi$($r_p, r_\pi$). We have highlighted the DD-$k$NN as it produces tighter constraints compared to the query-data $k$NN($r_p, r_\pi$).}
    \label{fig:corner_compare}
\end{figure*}

Figure~\ref{fig:corner_kNNs} showcases the parameter posteriors for the 3 different configurations of the $k$NN statistics. Green represents the standard 1D $k$NN($r$) statistic; red represents the 2D $k$NN($r_p, r_\pi$) statistic; blue represents the DD-$k$NN statistic. The black lines show the underlying truth values. The marginalized constraints shown at the top of the 1D PDF plots are the posterior mean and $2\sigma$ constraints of the DD-$k$NN statistic. We also summarize the marginalized constraints in Table~\ref{tab:constraints}. 

To quantify the difference in constraining power of the different summary statistics, we construct a Figure of Merit (FoM) metric as the following
\begin{equation}
    \mathrm{FoM} = \left[\mathrm{Vol}(<3\sigma)\right]^{-\frac{1}{n}},
    \label{equ:fom}
\end{equation}
where $n$ is the number of parameter in the model. $\mathrm{Vol}(<3\sigma)$ is the volume in the model parameter space of all samples within $3\sigma$ of the best-fit. We approximate this volume by computing the volume of the minimum enveloping ellipsoid of the $3\sigma$ samples. In this empirical scheme, the FoM roughly indicates the inverse of the constraint per parameter direction. The more informative statistics will have larger FoMs. We summarize the FoM values for the four summaries statistics in the bottom row of Table~\ref{tab:constraints}.

First of all, we get unbiased constraints on all parameters with each of the 3 $k$NN flavors. This adds support to the conclusions of our companion paper \citet{2022cYuan} and further confirms the efficacy of the lightcone-based full forward model approach for an additional set of statistics. We expect our forward model approach to also apply to other novel summary statistics. 

Figure~\ref{fig:corner_kNNs} also shows that, against the same mock galaxy catalog, and using the same $k$s and roughly the same scale cuts, the different formulations of the $k$NN statistics get drastically different constraints. Specifically, the $k$NN($r_p, r_\pi$) is able to derive much stronger constraints than the standard 1D $k$NN. Furthermore, the 2D $k$NN with data-data pairs is able to further tighten the constraints on the HOD parameters. It is likely that the 2D formulations get stronger constraints because they disentangle the redshift-space distortion effects whereas the standard $k$NN($r$) does not. Specifically, the line-of-sight structure on the small scales is strongly attributed to the velocity dispersion of the galaxies, which strongly depends on the 1-halo structure and the satellite fraction of the halos. Thus, capturing the RSD increases the constraining power on the HOD parameters. A similar effect was found for the 2PCF, where the full 2D $\xi(r_p, r_\pi)$ statistic was found to be significantly more constraining on HOD and assembly bias parameters than the projected 2PCF \citep{2021bYuan}.  

The fact that the DD-$k$NN derives stronger constraints on the HOD than the standard $k$NN($r_p, r_\pi$) is not surprising. The standard $k$NN tabulates distances to galaxy neighbors around randomly chosen query points. By definition, the query points will more likely populate low density regions instead of high density regions. Thus, the standard $k$NN up-weighs low density regions instead of high density regions. However, most of the HOD constraints come from high density regions, \red{as luminous galaxies tend to live in the more massive halos, and satellite occupation is better constrained in the more massive halos.} In the DD-$k$NN case, the query points are the galaxy positions, which populate dense regions more than underdense regions. Thus, the DD-$k$NN is expected to be much more sensitive to the HOD, thus yielding stronger constraints. Another way to understand the difference is that the DD-$k$NN is a mass-weighted re-average of the standard $k$NN, thus more sensitive to denser environments. 

However, the three different flavors of $k$NNs do result in comparable constraining power on parameter $\log M_\mathrm{cut}$ parameter. This parameter controls the central occupation in this model, and thus largely determines the amplitude of the galaxy power spectrum assuming a low satellite fraction. This suggests that all three flavors of $k$NNs can effectively capture the clustering amplitude. This in turn implies that the $k$NNs are potentially powerful probes for constraining cosmic growth rate, assuming we can disentangle galaxy bias from the underlying matter clustering amplitude. This is consistent with Fisher forecast results in \citet{2022Banerjee}.

Figure~\ref{fig:corner_compare} compares the parameter constraints from the 2D $k$NNs to the redshift-space 2PCF $\xi(r_p, r_\pi)$. The standard $k$NN($r_p, r_\pi$) delivers roughly equivalent constraining power on the HOD compared to the 2PCF. However, the DD-$k$NN derives considerably stronger constraints than the 2PCF across all parameters, especially on the mass parameters $\log M_\mathrm{cut}$ and $\log M_1$. This is consistent with the fact that the full shape 2PCF can be exactly derived from the DD-$k$NNs. 
\red{The fact that the $k$NN constraints and the 2PCF have different degeneracy directions in the posterior contours is consistent with the theoretical notion that the $k$NN taps into the information content of higher order correlation functions, but without the computational complexities. }

\red{The fact that DD-$k$NN and $k$NN($r_p, r_\pi$) have different degeneracies also suggests that the two statistics capture different features in the density field and that a joint analyses of these two statistics can be particularly fruitful. We leave that to future work as we do not have enough covariance realizations to confidently derive the joint covariance matrix between the two statistics.  }

\red{We list the FoMs of the four statistics in Table~\ref{tab:constraints}. It is also helpful to think of them as inverses of the sample size needed to achieve the same model constraints. Of couse, this is assuming that the constraining power is limited by sample variance, i.e. $\mathrm{FoM} \sim 1/\sqrt{\mathrm{Var}} \sim 1/\sqrt{N}$, where $N$ is sample size. This suggests that to achieve the same level of constraints on the HOD, the 2PCF requires 4 times the sample size as that of DD-$k$NN. This is significant given the high cost of modern surveys. Though it remains to be seen if such conclusions would extend to cosmological parameters. }

\red{These results clearly position the 2D $k$NNs as powerful tools in maximally capturing the information content of upcoming galaxy survey data across a wide range of scales. Specifically, DD-$k$NNs are highly sensitive probes of highly non-linear scales and should be particularly powerful in constraining and validating galaxy bias models. This also presents exciting opportunities in combining the 2D $k$NNs with other probes such as galaxy-galaxy lensing to break parameter degeneracies and test robustness of galaxy bias models. It could also be potentially interesting to combine traditional clustering probes with the query-data $k$NN($r_p, r_\pi$) as it accesses a different set of information in the density field that were not accessible via traditional clustering statistics. }



\subsection{Galaxy assembly bias}
\label{subsec:gab}

A commonly explored and physically motivated extension to the standard HOD model is the so-called galaxy assembly bias, i.e. galaxy occupation correlates with secondary halo properties beyond just halo mass \citep[e.g.][]{2018Wechsler, 2018Artale, 2018Zehavi, 2019Bose, 2019Contreras, 2020Hadzhiyska, 2020Xu, 2016Hearin, 2005Gao, 2005Zentner, 2006Zhu, 2006Wechsler, 2007Gao, 2007Croton, 2008Li}. Mathematically, it extends the standard HOD probability distribution $P(N_g|M_h)$ to $P(N_g|M_h, x)$, where $N_g$ is the number of galaxies, $M_h$ is the halo mass, and $x$ is some secondary halo property. Galaxy assembly bias is important because it not only alters the interpretation of the underlying galaxy evolution models, but also significantly impacts galaxy clustering and ignoring it can bias cosmology constraints \citep{2014Zentner, 2014Pujol, 2019Lange}. In this section, we compare the constraining power of the 2D $k$NN statistics and the full-shape 2PCF on galaxy assembly bias. 

Traditionally, galaxy concentration has been used as the secondary marker of galaxy secondary bias (also known as galaxy assembly bias) due to the concentration's correlation with halo age \citep[e.g.][]{2002Wechsler, 2007Croton, 2007Gao}, with older halos having higher concentrations and vice versa.
However, a series of recent studies found local environment of the halo to be an excellent tracer of galaxy secondary bias based on hydrodynamical simulations and semi-analytic models \citep[e.g.][]{2022Yuan, 2020Hadzhiyska, 2020Xu, 2021Xu, 2021Delgado}. Thus, for this test, we adopt the environment-based assembly bias parameterization ($B_\mathrm{cent}$, $B_\mathrm{sat}$) as implemented in \ahod. Briefly, the environment is defined as the mass density within a $r_\mathrm{env} = 5h^{-1}$Mpc tophat of the halo center, excluding the halo itself. $B_\mathrm{cent} = 0, B_\mathrm{sat} = 0$ indicate no environment-based secondary bias. A positive $B$ indicates a preference for halos in less dense environments, and vice versa. 

\begin{figure}
    \centering
    \hspace*{-0.3cm}
    \includegraphics[width = 3.4in]{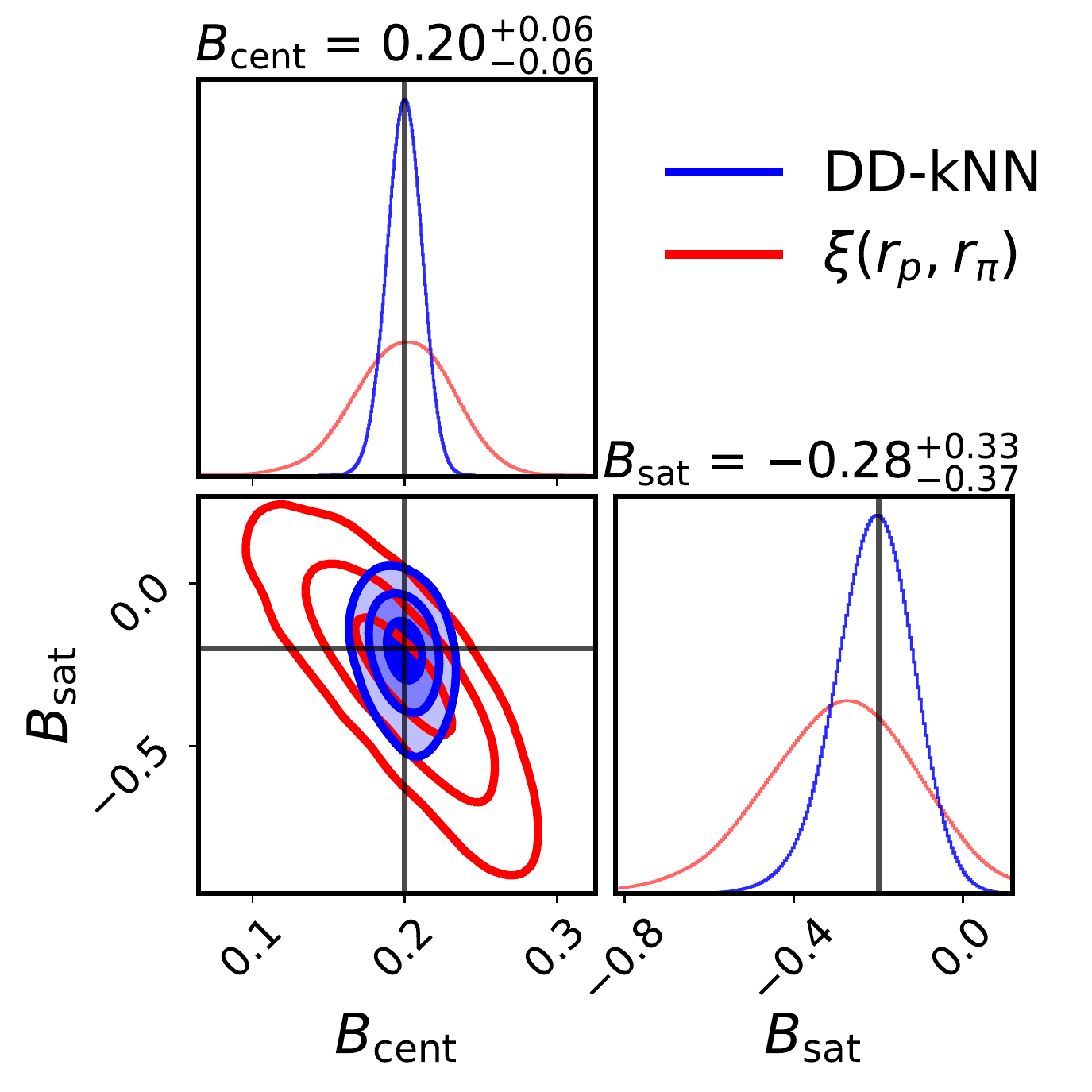}
    \vspace{-0.0cm}
    \caption{The marginalized recovery of galaxy assembly bias parameters $B_\mathrm{cent}$ and $B_\mathrm{sat}$. The black lines indicate the truth values in the mock catalog. The blue and red contours indicate the $1-3\sigma$ constraints with the DD-$k$NN and $\xi(r_p, r_\pi)$, respectively. The titles show the 95$\%$ constraints with $\xi(r_p, r_\pi)$. }
    \label{fig:corner_B}
\end{figure}

To evaluate the constraining power on $B_\mathrm{cent}$ and $B_\mathrm{sat}$ of our summary statistics, we follow the same procedure as described in section~\ref{sec:recovery}, but with a mock underlying HOD model that includes galaxy assembly bias, where $B_\mathrm{cent} = 0.2$ and $B_\mathrm{sat} = -0.2$. Physically, in this mock model the centrals prefer halos in denser environments whereas the satellites prefer halos in less dense environments. Figure~\ref{fig:corner_B} shows the marginalized posterior constraints on the two assembly bias parameters using the DD-$k$NN and the full shape 2PCF $\xi(r_p, r_\pi)$. \red{We summarize the marginalized posterior constraints in Table~\ref{tab:B}. While both summary statistics recover unbiased constraints on the assembly bias parameters, the DD-$k$NN clearly derives 2-3 times tighter constraint along either parameter direction, further demonstrating the information content of the 2D $k$NN statistics.} 

\begin{table}
\centering
\begin{tabular}{lccc}
\hline
\hline
Parameter   & truth  & $\xi(r_p, r_\pi)$ & DD-$k$NN$(r_p, r_\pi)$ \\
\hline
$B_\mathrm{cent}$  & 0.2 & 0.20$\pm 0.06$& $0.20\pm 0.02$\\
$B_\mathrm{sat}$   & -0.2 & $-0.28\pm 0.35$& $-0.21\pm 0.16$\\
\hline
\end{tabular}
\caption{The posterior constraints of the galaxy assembly bias parameters, using the redshift-space 2-point correlation function and the DD-$k$NN. The errorbars correspond to 95$\%$ uncertainties. }
\label{tab:B}
\end{table}

\section{Discussions and conclusions}
\label{sec:conclude}

Beyond 2PCF statistics are needed to summarize the rich information on small scales in the era of precision galaxy clustering measurements enabled by DESI and other upcoming surveys. \red{In this work, we introduce novel 2D generalizations of the $k$-th nearest neighbor statistics and use the lightcone-based forward model framework described in \citet{2022cYuan} to test the constraining power of these statistics on the galaxy--halo connection model. We show that the 2D $k$NNs readily generate other popular galaxy clustering statistics such as counts-in-cells, void probability functions, and the 2-point correlation function. Moreover, we show that the 2D $k$NN are computationally efficient. }

By conducting model recovery tests using the lightcone-based forward model, we find that the data-data 2D $k$NN formulation extracts the most galaxy--halo connection information among the variations. We compare the DD-$k$NN to the full-shape 2PCF $\xi(r_p, r_\pi)$ and find that the DD-$k$NN is a significantly stronger probe of essentially all aspects of the galaxy--halo connection model, including galaxy assembly bias. This is consistent with the theoretical notion that the $k$NNs encode not just the 2-point information, but also information from higher order $n$-point statistics. These tests also demonstrate the additional statistical power accessible on small scales through novel summary statistics. Our forecasts are also robust through the use of the lightcone-based forward model, exposing the summary statistics to the full range of observational systematics and survey realism. 

\red{The goal of developing these novel statistics is to be able to access and disentangle the galaxy--halo connection and cosmology information on non-linear scales. In the next paper, we will extend our model to include cosmology variations. So far, the tight constraints we get on $M_\mathrm{cut}$ at fixed cosmology suggests that the 2D $k$NNs tightly constrain the amplitude of galaxy clustering. This means that if we can break the degeneracies between galaxy bias and the amplitude of the matter power spectrum (for example by combining DD-$k$NN with galaxy-galaxy lensing), the 2D $k$NNs can prove to be a powerful probe of cosmology. This is particularly fruitful and timely as DESI and other survey facilities come online, which will produce high-density high-completeness samples particularly suited for small-scale analyses.}

\section*{Acknowledgements}

We would like to thank Risa Wechsler, Arka Banerjee, Ashley Ross, Sebastian Wagner-Carena, Philip Mansfield, and others for useful feedback and suggestions in various stages of this analysis.

This work was supported by U.S. Department of Energy through grant DE-SC0013718 and 
under DE-AC02-76SF00515 to SLAC National Accelerator Laboratory. 

This work used resources of the National Energy Research Scientific Computing Center (NERSC), a U.S. Department of Energy Office of Science User Facility located at Lawrence Berkeley National Laboratory, operated under Contract No. DE-AC02-05CH11231.

The {\sc AbacusSummit} simulations were conducted at the Oak Ridge Leadership Computing Facility, which is a DOE Office of Science User Facility supported under Contract DE-AC05-00OR22725, through support from projects AST135 and AST145, the latter through the Department of Energy ALCC program.

\section*{Data Availability}

The simulation data are available at \url{https://abacussummit.readthedocs.io/en/latest/}. The \ahod\ code package is publicly available as a part of the \textsc{abacusutils} package at \url{http://https://github.com/abacusorg/abacusutils}. Example usage can be found at \url{https://abacusutils.readthedocs.io/en/latest/hod.html}.



\bibliographystyle{mnras}
\bibliography{biblio} 







\bsp	
\label{lastpage}
\end{document}